\documentclass[a4paper,fleqn,twocolumn,10pt]{article}

\usepackage{bm,amsmath,amsthm,hyperref,amsfonts,amssymb,color}
\usepackage{mathrsfs} 
\usepackage{cite}

\newcommand\pdt[1]{\frac{\partial{#1}}{\partial t}} 
\newcommand{\lineunder}[2]{\LU{\begin{array}[t]{c}\underbrace{#1}\vspace*{.5em}\end{array}}{\mbox{\footnotesize\rm #2}}}
\newcommand{\LU}[2]{\begin{array}[t]{c}#1\vspace*{-1em}\\_{#2}\end{array}}
\newcommand{\linesunder}[3]{\LSU{\begin{array}[t]{c}\underbrace{#1}\vspace*{.5em}\end{array}}{\mbox{\footnotesize\rm #2}}{\mbox{\footnotesize\rm #3}}}
\newcommand{\LSU}[3]{\begin{array}[t]{c}#1\vspace*{-1em}\\_{#2}\vspace*{-.5em}\\_{#3}\end{array}}
\newcommand{\morelinesunder}[4]{\LSUU{\begin{array}[t]{c}\underbrace{#1}\vspace*{.5em}\end{array}}{\mbox{\footnotesize\rm #2}}{\mbox{\footnotesize\rm #3}}{\mbox{\footnotesize\rm #4}}}
\newcommand{\LSUU}[4]{\begin{array}[t]{c}#1\vspace*{-1em}\\_{#2}\vspace*{-.5em}\\_{#3}\vspace*{-.5em}\\_{#4}\end{array}}

\usepackage{mathrsfs}   
\usepackage{eucal}      
\usepackage{upgreek}    

\def\bbD{{\mathbb D}}  
  \def\bbI{{\mathbb I}}

\def\FG{\boldsymbol}
   
\def\dd{{\FG d}} \def\ee{{\FG e}} \def\ff{{\FG f}} 
\def\gg{{\FG g}} \def\hh{{\FG h}} 
\def\jj{{\FG j}}   
\def\mm{{\FG m}} \def\nn{{\FG n}}  
  \def\rr{{\FG r}} 
\def\tt{{\FG t}} \def\uu{{\FG u}} 
\def\vv{{\FG v}}  \def\xx{{\FG x}} 
\def\yy{{\FG y}}  
   
\def\DD{{\FG D}} 
\def\FF{{\FG F}} 

 \def\KK{{\FG K}} \def\LL{{\FG L}}

\def\SS{{\FG S}} \def\TT{{\FG T}} 
  \def\XX{{\FG X}}

\newcommand{\R}{\mathbb R}

\newcommand{\Nabla}{{\nabla}}
\newcommand{\Fe}{\FF_\text{\!\rm e\,}}
\newcommand{\Fp}{\FF_\text{\!\rm i\,}}
\newcommand{\Fpp}{\FF_\text{\!\rm i\,}}
\newcommand{\Fetop}{\FF_\text{\!\rm e\,}^{\top}}
\newcommand{\Lp}{\LL}

\newcommand{\pl}{\partial}
\newcommand{\eq}[1]{(\ref{#1})}
\renewcommand{\d}{\mathrm d}  

\newcommand{\bulet}{\text{\footnotesize$\,\bullet\,$}}
\newcommand\ZJ[1]{\mathchoice
                 {{\buildrel{\hspace*{.1em}{_{_{\Large\boldsymbol\circ}}}}\over{#1}}}
                 {{\buildrel{\hspace*{.1em}{_{_{\Large\boldsymbol\circ}}}}\over{#1}}}
                 {{\buildrel{\hspace*{.1em}{_{_{\Large\boldsymbol\circ}}}}\over{#1}}}
                 {{\buildrel{\hspace*{.1em}{\boldsymbol\circ}}\over{#1}}}}
\newcommand\DT[1]{\mathchoice
                 {{\buildrel{\hspace*{.1em}\text{\LARGE.}}\over{#1}}}
                 {{\buildrel{\hspace*{.1em}\text{\LARGE.}}\over{#1}}}
                 {{\buildrel{\hspace*{.1em}\text{\Large.}}\over{#1}}}
                 {{\buildrel{\hspace*{.1em}\text{\large.}}\over{#1}}}}
\def\W{w}
\def\OMEGA{\omega}
\def\DIS{\bbD}
\def\COND{k}
\def\HC{h_\text{\rm c}}
\def\GM{G_\text{\sc m}^{}}

\newtheorem{theorem}{Theorem}[section]

\newtheorem{example}[theorem]{Example}

\newtheorem{remark}[theorem]{Remark}

\numberwithin{equation}{section}

\usepackage{graphicx}
\usepackage{psfrag} 
\newcounter{myfigure}
\newenvironment{my-picture}[3]{\refstepcounter{myfigure}\label{#3}\setlength{\unitlength}{1cm}\begin{picture}(#1,#2)}{\end{picture}}

\definecolor{labelkey}{rgb}{1.,.2,0.}

\def\GRAVITY{\gg}
\def\COUPLING{\phi}
\def\bmxi{\boldsymbol\xi}

\usepackage{authblk}

\begin{document}

\title{{\bf Thermomechanics of ferri-antiferromagnetic phase
  transition in finitely-strained rocks\\towards paleomagnetism}}

\author[$1$,$2$]{Tom\'a\v s Roub\'\i\v{c}ek}

\affil[$1$]{Mathematical Institute, Charles University,
            Sokolovsk\'a 83, CZ-186~75~Praha~8, Czechia}

\affil[$2$]{Institute of Thermomechanics, Czech Acad.\ Sci.,
          Dolej\v skova 5, CZ-18200~Praha~8, Czechia}

\twocolumn[
  \begin{@twocolumnfalse}
    \maketitle

\begin{abstract}
The thermodynamic model of visco-elastic deformable magnetic materials
at finite strains is formulated in a fully Eulerian way in rates with the
aim to describe thermoremanent paleomagnetism in crustal rocks.
The Landau theory applied to a ferro-to-para-magnetic phase transition,
the gradient theory for magnetization (leading to exchange energy) with
general mechanically dependent coefficient, hysteresis in magnetization
evolution by Gilbert equation involving objective corotational
time derivative of magnetization, and demagnetizing field are considered in
the model. The Jeffreys viscoelastic rheology
is used with temperature-dependent creep to model solidification or melting
transition. The model complies with energy conservation and the
Clausius-Duhem entropy inequality.

\bigskip


\end{abstract}

\bigskip\bigskip\bigskip\bigskip

\end{@twocolumnfalse}
]

\section{Introduction}\label{sec-intro}

Deformable magnetic media are an interesting multi-physi\-cal
area of continuum mechanics. Applications to paleomagnetism
in crustal rocks is particularly interesting because
it combines sophisticated viscoelastic rheology with
thermomechanics and with mechanical and magnetical phase transitions,
i.e.\ liquid magma to essentially solid rocks and para- (or rather
antiferro-) magnetism to ferro- (or rather ferri-) magnetism
in rocks. Magnetism in (some) rocks forms a vital part of rock
physics and mechanics, cf.\
\cite{Butl92PMDG,Camp03IGF,DunOzd97RM,LanMel06EM,StaBan74PPRM}.
Paleomagnetism refers to ``frozen'' magnetism in
oceanic or continental rocks, which may give information about
history of geomagnetic field generated in the Earth's outer core or
history of deformation of continental crust, respectively. Interestingly,
paleomagnetism exists also in other planets that nowadays
do not have substantial magnetic fields, specifically Mars and Mercury.
Cf.\ \cite{Butl92PMDG} and
\cite{CoxDoe60RP} for a survey.

\begin{figure}[]
	\centering
	\includegraphics[width=.47\textwidth]{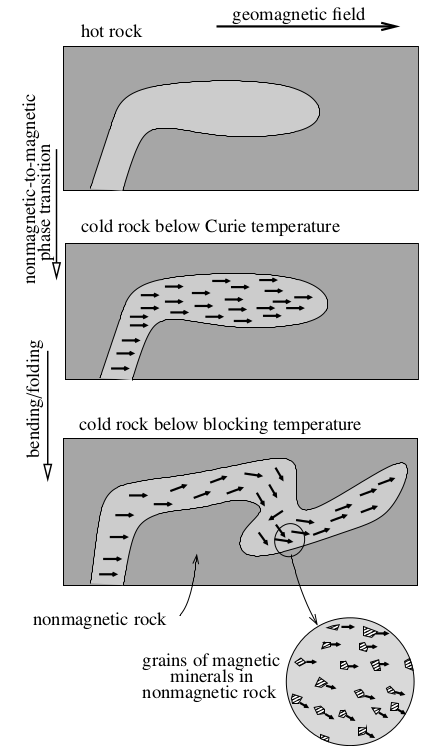}
 \caption{{\sl A schematic illustration of thermoremanent magnetization within
          cooling and deformation of magnetic rocks.}
        }\label{fig1}
\end{figure}

Some chemical components in rocks as iron oxides (as magnetite and
hematite) and some other oxides (e.g.\ iron-titanium oxide in basalts)
are ferrimagnetic in low temperatures. They form a single- or poly-crystalic
grains in mostly nonmagnetic silicate rocks and can be magnetized by various
mechanisms: Most important (which we are focused on) is 
{\it thermoremanent magnetization} in so-called igneous rocks which are
formed through the cooling and solidification of magma or lava. 
and subsequent bending/folding of rocks in a constant geomagnetic field.
The processes within gradual
cooling and magnetization and subsequent deformation of rocks within
long geological timescales are schematically depicted in  Fig.\,\ref{fig1};
in fact, grains of magnetic minerals have randomly oriented
easy-mag\-netization axes, some of them being magnetized more while others
less. The other  mechanisms are detrital remanent magnetization (in sediments),
isothermal remanent magnetization (in strong magnetic fields typically
during lightning at a fixed temperature),  etc.

The deformation of rocks within long-time scales can surely be
very large, both in the oceanic crust and in the continental crust, too.
Thus large-strain setting is to be used. Here we will exploit
the fully nonlinear continuum mechanics of magnetic materials as
devised for isothermal situations by \cite{DeSPoG95ISIS,DeSPoG96CTDF}.
Together with the anisothermal Landau phase-transition theory, applied
for rigid magnets as by \cite{PGRoTo10TCTF}, it gives a full thermomechanical
model of deformable magnetic continua in the solid-type Kelvin-Voigt rheology,
as analyzed in a multipolar variant by \cite{Roub23LTFP}. This is here
presented in Section~\ref{sec-elast-mag}. To model the fluidic character of
hot rocks (magma) and long-time-scale deforming cold rocks and the
solidification phase transition, we must use a suitable rheology of the
Maxwell type. This combination is formulated in Section~\ref{sec-creep},
together with specifying the energetics behind the system and notes about
analytical justification in a multipolar variant involving higher-order
dissipative terms. Eventually, in Section~\ref{sec-rocks}, the application
to paleomagnetism is briefly specified.

The main notation used below is summarized in Table~1.

\begin{table}[]
\caption{Nomenclature}\label{tbl1}

  \begin{tabular*}{.4\textwidth}{@{\extracolsep{\fill}}|l|l|}

 $\vv$ velocity (in m/s),  & $\GRAVITY$\,gravity acceleration\,(in\,m/s$^2$),\\ 
$\varrho$ mass density (in kg/m$^3$), & $\ee(\vv)$
small strain rate (in s$^{-1}$),\\

$\FF$ deformation gradient,& $\DIS$ viscosity-coefficient tensor,\\

$\Fe$\,elastic distortion,& $\psi$ free energy (in Pa=J/m$^3$),\\

$\Fp$\,inelastic (creep) distortion,\!\!\!& $\zeta$\,dissipation potential\,(in\,Pa/s),\!\!\\

$\Lp$ inelastic distortion rate,& $c$ heat capacity (in Pa/K),\\

$\mm$ magnetization (in A/m),& $\COND$\,heat conductivity,
\\

$\theta$ temperature (in K),&  $\kappa$ exchange coefficient,\\

$u$ demagnetizing potential, & $\eta$ entropy (in Pa/K),\\

$\mu_0$ vaccum permeability, &$\hh$\,total magnetic field\,(in\,A/m), \\

$\TT$\,elastic Cauchy stress, 
& $\:\hh_{\rm ext}$ external magnetic field,\\

$\KK$\,Korteweg-like stress, & $\:\gamma$ gyromagnetic ratio,\\

$\SS$ magnetic-dipole stress,& $(^{_{_{\bullet}}})\!\DT{^{}}
$ convective time derivative,\\

$\mathscr{S}$ couple stress (in Pa\,m), &
$\!(^{_{_{\bullet}}})\!\ZJ{^{}}$corotational time derivative.\!\!\\

\end{tabular*}
\end{table}

\section{Viscoelastic finitely\\strained magnets}\label{sec-elast-mag}

The basic bricks for building our magneto-thermo-viscoelastic model are
continuum mechanics, Landau's theory of phase transition applied to magnetism
in deforming media, and thermomechanics.

\subsection{Eulerian continuum mechanics}\label{sec-elast-mag1}

The basic kinematic concept is the
time-evolving deformation $\yy:\varOmega\to\R^3$ as a mapping from a reference
configuration of the body $\varOmega\subset\R^3$ into a physical space $\R^3$.
The ``Lagrangian'' space variable in the reference configuration will be
denoted as $\XX\in\varOmega$ while in the ``Eulerian'' physical-space
variable by $\xx\in\R^3$. The basic kinematic and geometrical objects are
the Lagrangian velocity $\vv=\pdt{}\yy$ and the Lagrangian
deformation gradient $
\Nabla_{\!\XX}^{}\yy$.

Time evolving deformations $\xx=\yy(t,\XX)$ are sometimes called ``motions''.
Further, assuming for a moment that $\yy(t,\cdot)$ is invertible, we define
the so-called {\it return} (sometimes called also a {\it reference})
{\it mapping} $\bmxi:\xx\mapsto\yy^{-1}(t,\xx)$. The important quantities are
the Eulerian velocity $\vv(t,\xx)=\vv(t,\bmxi(t,\xx))$ and the Eulerian
deformation gradient $\FF(t,\xx)=[\Nabla_{\!\XX}^{}\yy](t,\bmxi(t,\xx))$.
We use the dot-notation
$(\cdot)\!\DT{^{}}=\pdt{}+\vv{\cdot}\nabla_\xx$ for the {\it convective time
derivative} applied to scalars or, component-wise, to vectors or tensors.
Then the velocity gradient
$\Nabla\vv=\nabla_{\!\XX}^{}\vv\nabla_{\!\xx}^{}\XX=\DT\FF\FF^{-1}$,
where we used the chain-rule calculus  and
$\FF^{-1}=(\nabla_{\!\XX}^{}\xx)^{-1}=\nabla_{\!\xx}^{}\XX$. 
This gives the {\it transport equation-and-evolution  for the
deformation gradient} as
\begin{align}
\DT\FF=(\nabla\vv)\FF\,.
\label{ultimate}\end{align}
The return mapping $\bmxi$ satisfies the transport equation
\begin{align}
\DT{\bmxi}=\boldsymbol0\,;
\label{transport-xi}\end{align}
note that, since we confined ourselves to a spatially homogeneous material
(except Remark~\ref{rem-hetero} below), 
actually $\bmxi$ does not explicitly occur in the formulation of
the problem, although we could use it while substituting
$\FF=(\nabla\bmxi)^{-1}$. 

\subsection{Free energy and dissipation potential}

Beside the mechanical variables in Section~\ref{sec-elast-mag1}, we
consider the magnetization vector field $\mm$ and temperature $\theta$.
The main ingredients of the model are the (volumetric) {\it free energy}
$\psi=\psi(\FF,\mm,\theta)$ considered per the {\it referential volume}
 (so at this point $\FF$ and $\mm$ and $\theta$ are meant only
as placeholders where later the actually fields are placed) 
and mechanical and magnetic  {\it dissipative-forces}.
 It should be noted that such free energy (considered in J/m$^3$=Pa)
correspond to ``standard'' physical data while the actual
free energy (considered also in J/m$^3$=Pa) is 
$\psi(\FF,\mm,\theta)/\!\det\FF$. Let us remark that an alternative
referential energy $\psi$ could be (and sometimes is) considered
in J/kg so that than the actual energy wold be $\varrho\psi$. In any case,
the actual free energy acts on the actual (i.e.\ Eulerian) 
deformation gradient $\FF$, actual magnetization $\mm$, and actual
temperature $\theta$ as functions of the Eulerian $\xx$ and of time $t$.

In addition, it is conventional in micromagnetism
(cf.\ e.g.\ \cite{Brow66MI}) to augment the free energy
also by an exchange energy $\upkappa(\FF)|\nabla\mm|^2$.
The free energy considered per actual (not referential) volume extended
by the Zeeman energy arising by an applied external actual (not referential)
magnetic field $\hh_{\rm ext}$, i.e.\ the Gibbs-type {\it actual free energy},
is thus
\begin{align}\nonumber
\hspace*{-2em}\psi_\text{\sc g}^{}(t;\FF,\mm,\nabla\mm,\theta)&=
\!\!\!\!\!
\lineunder{\frac{\psi(\FF,\mm,\theta)}{\det\FF}}{actual free energy}
\!\!\!\!\!\!
\\[-.5em]&\hspace*{.7em}-\!\!\!\!\!\!
\linesunder{\mu_0\hh_{\rm ext}(t){\cdot}\mm_{_{_{}}}}{Zeeman}{energy}
\!\!\!\!\!\!\!+\!\!\!\!\!\!\linesunder{\frac{\upkappa(\FF)|\nabla\mm|^2}{2
\det\FF}}{exchange}{energy}\!\!\!\!\!
\label{ansatz}\end{align}
with the coefficient $\upkappa$ depending generally on $\FF$.

From the free energy \eq{ansatz}, we can read as partial (functional)
derivatives of $\psi$ with respect to $\FF$, $\nabla\mm$, $\mm$, and $\theta$
respectively the {\it conservative part of the Cauchy stress} $\TT$,
a {\it capillarity}-like {\it stress} $\KK$, the actual conservative
{\it magnetic driving force} $\tt$, and the {\it entropy} $\eta$ as:
\begin{subequations}\label{driving-fields}\begin{align}\nonumber
    &\hspace*{-2em}\TT=\frac{
[\psi_{\rm G}^{}]_\FF'(\FF,\mm,\nabla\mm,\theta)\FF^\top\!\!\!}{\det\FF}
\\[-.7em]&\label{driving-stress}=
\bigg(\!
\psi_\FF'(\FF,\mm,\theta)
+\frac{\upkappa'(\FF)|\nabla\mm|^2\!}2\;\bigg)\frac{\FF^\top}{\det\FF}\,,
\\
&\hspace*{-2em}\KK=\frac{(\nabla\mm)^\top\psi_{\nabla\mm}'(\FF,\nabla\mm)}{\det\FF}=
\mu_0\upkappa(\FF)\frac{\nabla\mm{\otimes}\nabla\mm}{\det\FF}\,,
\label{capillarity-stress}
\\\nonumber
&\hspace*{-2em}\tt=\frac{[\psi_{\rm G}^{}]_\mm'(\FF,\mm,\theta)}{\det\FF}-{\rm div}\frac{[\psi_{\rm G}^{}]_{\nabla\mm}'(\FF,\nabla\mm)}{\det\FF}-\mu_0\hh_{\rm ext}
\\&\hspace*{-2em}\ \ 
=\frac{
  \psi_\mm'(\FF,\mm,\theta)\!\!}{\det\FF}
-{\rm div}\Big(\frac{\upkappa(\FF)\nabla\mm}{\det\FF}\Big)-\mu_0\hh_{\rm ext}
\,,\ \text{ and}
\label{driving-field}\\
&\hspace*{-2em}\eta=-\frac{\psi_\theta'(\FF,\mm,\theta)}{\det\FF}
\,.
\label{stress-entropy}\end{align}\end{subequations}
The product $\nabla\mm{\otimes}\nabla\mm$ in \eq{capillarity-stress} is to be
understood componentwise, specifically  $[\nabla\mm{\otimes}\nabla\mm]_{ij}$
$=\sum_{k=1}^3\frac{\pl}{\pl x_i}m_k\frac{\pl}{\pl x_j}m_k$ with
$\mm=(m_1,m_2,m_3)$ and $\xx=(x_1,x_2,x_3)$.

The other mentioned ingredient of our model is dissipative forces. It is
conventional (although not necessary) to read them from the mentioned
dissipative-force potential $\zeta=\zeta(\theta;\ee,\rr)$ with
$\ee=\ee(\vv)=\frac12\Nabla\vv^\top\!+\frac12\Nabla\vv$ and with $\rr$
the magnetization rate to be specified later in \eq{def-ZJ}.
This determines the mechanical dissipative stress
$\DD=\zeta_{\ee}'(\theta;\ee,\rr)$ and the magnetic dissipative force
$\dd=\zeta_{\rr}'(\theta;\ee,\rr)$.

The {\it momentum equilibrium} equation then balances the divergence of
the total Cauchy stress with the inertial and gravity force:
\begin{align}
\hspace*{-1em}\varrho\DT\vv-{\rm div}\big(\TT{+}\DD{+}\TT_{\!\rm mag}^{}
        {-}\,{\rm div}\,\mathscr{S}
\big)=\varrho\GRAVITY+\ff_{\!\rm mag}^{}
\label{Euler-thermodynam1-}\end{align}
with $\TT$ from \eq{stress-entropy}.
Moreover, $\TT_{\!\rm mag}^{}$ and $\ff_{\!\rm mag}^{}$
are the magnetic stress and the magnetic force which balance the
energetics, specifically
\begin{align*}\hspace*{-1.5em}
  \TT_{\!\rm mag}^{}\!\!:=\KK+\SS\ \ \text{ and }\ \
\ff_{\!\rm mag}^{}\!\!:=\mu_0(\nabla\hh)^\top\mm-\mu_0\nabla(\hh{\cdot}\mm)
\end{align*}
where $\SS$ is the skew-symmetric magnetic-dipole stress
while $\mathscr{S}$ will be a ``magnetic exchange hyperstress'',
 specifically
\begin{align*}
&\hspace*{-1.5em}\SS={\rm skw}\big((\mu_0\hh {-}
\psi_\mm'(\FF,\mm,\theta)/\!\det\FF){\otimes}\mm\big)\ \ \ \text{ and}
\\&\hspace*{-1.5em}\mathscr{S}=
{\rm Skw}\big(\mm{\otimes}[\psi_{\rm G}^{}]_{\nabla\mm}'(\FF,\nabla\mm)\big)
=\frac{\!\upkappa(\FF){\rm Skw}(\mm{\otimes}\nabla\mm)\!}{\det\FF}\,,
\end{align*}
where the skew-symmetric part ``Skw''
of the 3rd-order tensor is defined as
\begin{align}
\big[{\rm Skw}(\mm{\otimes}\nabla\mm)\big]_{ijk}
:=\frac12\Big(m_i\frac{\pl m_j}{\pl x_k}-m_j\frac{\pl m_i}{\pl x_k}\Big)\,.
\end{align}

The driving magnetic force $\tt$ in \eq{driving-field} enters the
Gilbert equation \eq{LL-friction} in the next section
while the entropy $\eta$ in \eq{stress-entropy} will be the departure point
for the formulation of the heat equation in Sect.\,\ref{sec-therm}.

\subsection{Landau theory of magnetic phase transition}

L.D.\,\cite{Land37TPT} devised a pioneering theory of phase
transitions. The essence is in a simple polynomial free energy
that changes its convex-vs-nonconvex character smoothly within
varying temperature. Here, the free energy being 4th-order polynomial
in terms of the magnetization $\mm$ reads as
\begin{align}\label{Landau-ansatz}
\hspace*{-2em}\psi=\psi(\mm,\theta)=a|\mm|^4\!+b(\theta{-}\theta_\text{\sc c}^{})|\mm|^2\!
  +c\theta(1{-}{\rm ln}\theta)
\end{align}
with $\theta_\text{\sc c}^{}>0$ the Curie (or Ne\'el) transition temperature,
$a,b>0$, and $c>0$ the heat capacity. Note that the function
$\psi=\psi(\cdot,\theta)$ is convex for $\theta\ge\theta_\text{\sc c}^{}$,
while for $\theta<\theta_\text{\sc c}^{}$ it is nonconvex.
In static magnetically soft magnets, the magnetization minimizes the
energy. Here the minimum of $\psi(\bulet,\theta)$ is attained on the orbit
$|\mm|=m_\text{\sc s}(\theta)$ with the radius  (called {\it saturation
 } or {\it spontaneous magnetization}) 
\begin{align}\label{ms}
  m_\text{\sc s}(\theta)=
  \begin{cases}\sqrt{a(\theta_\text{\sc c}{-}\theta)/(2b)}&\text{if }\
    0\le\theta\le\theta_\text{\sc c},\\ 0 &\text{if }\
    \theta\ge\theta_\text{\sc c}\,,\end{cases}
\end{align}
cf.\ the middle line in Figure~\ref{fig3} below. Noteworthy, in the external
magnetic field $\hh_{\rm ext}$, the contribution in \eq{ansatz}
leads to the (slight) violation of the so-called Heisenberg constraint
$|\mm|=m_\text{\sc s}(\theta)$, cf.\ Fig.\,5.4 in \cite{Bert98HM}.
This constraint is often considered non-realistically in mathematical
literature dealing with isothermal ferromagnetic modelling and,
among other drawbacks, would not allow for anisothermal
extension.\footnote{Even in a convexified (relaxed) variant in a
rigid magnets, the attempt by \cite{Roub03MFSS} for an anisothermal
extension with the Heisenberg constraint is extremely cumbersome.}

In time-dependent 
situations, employing a magnetization rate $\rr$, the evolution of $\mm$ is
conventionally governed by the {\it Gilbert equation} 
$\rr/\gamma=\mm{\times}(\mu_0\hh{-}\tt{-}\alpha\rr)$
with $\alpha>0$ a viscous-like damping constant, $\gamma=\gamma(\mm,\theta)$
a gyromagnetic ratio, $\hh$ an effective magnetic field, and $\tt$ a
magnetic driving force from \eq{driving-field}. This
can also be written in the Landau-Lifschitz form as
$\rr+\gamma\mm{\times}(\mu_0\hh{-}\tt)
=-\lambda\mm{\times}(\mm{\times}(\mu_0\hh{-}\tt))$ with a suitable $\lambda$.
Assuming for a moment $|\mm|$ constant, the Gilbert
equation can be rewritten into a more convenient form 
$\alpha\rr
-(\mm{\times}\rr)/\gamma=\mu_0\hh-\tt$, cf.\ \cite{RoToZa09GEDF}.

This simple linear damping corresponds to a quadratic dissipation
potential $\zeta(\rr)=\frac12\alpha|\rr|^2$ with $|\cdot|$ denoting
the Euclidean norm on $\R^3$, reflecting that we have in mind an
isotropic situation in polycrystalline magnetic rocks.
In many applications and in particular in paleomagnetism,
the magnetic evolution is an activated process due to pinning effects
which need certain activation energy for movement micro-magnetic walls.
This can be described by adding a dry-friction-type
1-homogeneous nonsmooth term into the dissipation potential
$\zeta(\rr)=\frac12\alpha|\rr|^2+\HC|\rr|$ with $\HC=\HC(\theta)$
a so-called coercive force. The magnetic dissipative force
is then $\dd=\zeta'(\rr)=\alpha\rr+\HC{\rm Dir}(\rr)$
with ``Dir'' denoting the set-valued monotone ``direction'' mapping
\begin{align}
{\rm Dir}(\rr)=\begin{cases}\{\rr\in\R^{3};\ |\rr|\le1\}&\text{if }\rr={\boldsymbol0}\,,\\
\quad\rr/|\rr|&\text{if }\rr\ne{\boldsymbol0}\,,\end{cases}
\label{LLG}\end{align}
cf.\ \cite{RoToZa09GEDF}; let us note the corresponding dissipation rate
is $\dd{\cdot}\rr=\alpha|\rr|^2+\HC|\rr|$. This nonsmooth extension was
proposed by \cite{BalHel91DFM} and \cite{Visi97MLLE} as a device to model
properly a {\it hysteretic response} in magnetization of ferromagnets,
modifying the Gilbert equation by augmenting suitably the effective magnetic
field. Although the original 
\cite{Gilb55LFGE} and the \cite{LanLif35TDMP}
equations are equivalent to each other, the resulting augmented equations
are no longer mutually equivalent. This has been pointed out
in \cite{PoGu01DMM}, where the conceptual differences between the Gilbert and
the Landau-Lifschitz formats have been elucidated.

In rigid magnets, simply $\rr=\pdt{}\mm$. Yet, in deforming media in Eulerian
description, the partial time derivative $\pdt{}\mm$  
should be replaced by an objective time derivative. Here
we use the Zaremba-Jaumann (corotational) time derivative $\ZJ\mm$, defined as
\begin{align}
  \hspace*{-1.7em}\ZJ\mm=\DT\mm-{\rm skw}(\nabla\vv)\mm
\ \ \ \text{ with }\ \ \
  \DT\mm=\pdt\mm+(\vv{\cdot}\nabla)\mm\,.
\label{def-ZJ}\end{align}
Thus, for $\rr=\ZJ\mm$, the Gilbert equation with dry friction turns into
\begin{align}\label{LL-friction}
\alpha\ZJ\mm+\HC(\theta){\rm Dir}(\ZJ\mm)
-\frac{\mm{\times}\ZJ\mm}{\gamma(\mm,\theta)}\ni\mu_0\hh-\tt\,,
\end{align}
the inclusion ``$\ni$'' being related to that the left-hand side
is set-valued at 
the zero rate. Moreover, in a deforming continuum, we can 
consider a more general $\FF$-dependent
$\gamma=\gamma(\FF,\mm,\theta)$ and  $\HC=\HC(\FF,\theta)$ but, rather due to
notational simplicity, we will not explicitly consider it.

Let us emphasize that the convective derivative $\DT\mm$ itself is not
objective and would not be suitable in our context, except perhaps some
laminar-like deformation as implicitly used in an incompressible isothermal
variant by 
\cite{BFLS18EWSE} or \cite{Zhao18LWPB} or in a nanoparticle transport in
fluids by \cite{GruWei21FITM}.

In deformable (and deforming) magnetic media, the Zaremba-Jaumann corotational
derivative for magnetization was suggested already by 
\cite{Maug76CTDF} to model situations when the magnetization can
be ``frozen'' in hard-magnetic materials in their ferro- or ferri-magnetic
state. For this effect, it is important the left-hand side in  
\eq{LL-friction} contains the function ``Dir'' which is set-valued
at $\ZJ\mm=0$ so that for $h_{\rm c}$ large (which will occur below the
so-called blocking temperature as depicted in Fig.\,\ref{fig3} below),
necessarily $\ZJ\mm=0$ so that $\mm$ exhibits the mentioned
``frozen'' effect. Later, the Zaremba-Jaumann derivative was used in
\cite{DeSPoG95ISIS,DeSPoG96CTDF} in the linear viscosity (magnetic attenuation)
$\alpha$-term.

The total magnetic field $\hh$ in \eq{LL-friction} is a difference of
an external (given) magnetic field $\hh_{\rm ext}^{}$ and the {\it demagnetizing
field} $\hh_{\rm dem}^{}$ self-induced by the magnetization itself.
For geophysical applications in Sect.\,\ref{sec-rocks}, the
full Maxwell electromagnetic system is considered simplified to
{\it magnetostatics}, considering slow evolution and neglecting
in particular eddy currents and even confining ourselves on electrically
non-conductive media. Then $\hh_{\rm dem}=-\nabla u$ with $u$ denoting
a scalar-valued potential solving the Poisson-type equation
\begin{align}\label{u-eq}
{\rm div}(\nabla u-\chi_{\varOmega}^{}\mm)=0
\end{align}
considered (in the sense of distribution) on the whole Universe with
$\chi_{\varOmega}^{}=\chi_{\varOmega}^{}(\xx)=1$ on $\varOmega$ while $=0$ outside
$\varOmega$. Fixing $u(\infty)=0$, in our 3-dimensional case, there is
the explicit integral formula for $u$, see \eq{Euler-thermodynam5} below.

\subsection{Thermodynamics}\label{sec-therm}

The further ingredient of the model is the {\it entropy equation} for
the entropy $\eta$ from \eq{stress-entropy}:
\begin{align}
\hspace*{-2em}\pdt\eta+{\rm div}\big(\vv\,\eta\big)
  =\frac{\xi-{\rm div}\,\jj}\theta\ \ \ \ \text{ with }\ 
  \jj=-\COND\nabla\theta \,
\label{entropy-eq}\end{align}
and with $\xi=\xi(\FF,\theta;\ee(\vv),\ZJ\mm)$
denoting the heat production rate specified later in
\eq{Euler-thermodynam3} and $\jj$ the heat flux governed by the
Fourier law with the thermal conductivity $\COND=\COND(\FF,\theta)$.
In the thermo-mechanically isolated system with $\vv{\cdot}\nn=0$ and
$\nabla\theta{\cdot}\nn=0$ on the boundary $\varGamma$ of $\varOmega$,
integrating \eq{entropy-eq} over $\varOmega$ and using Green formula
gives the {\it Clausius-Duhem inequality}, i.e. 
\begin{align}
  \frac{\d}{\d t}\int_\varOmega\eta\,\d\xx
=\int_\varOmega\!\!\!\!\!\!\!\!\!\lineunder{\frac\xi\theta+\COND\frac{|\nabla\theta|^2}{\theta^2}}{entropy production rate}\!\!\!\!\!\!\!\!\!\!\!\d\xx
\ge
0\,.
\label{entropy-ineq}\end{align}
i.e.\ \eq{entropy-eq} ensures the {\it 2nd law of thermodynamics}, saying
that the total entropy in isolated systems is nondecreasing in time.
Substituting $\eta$ from \eq{stress-entropy} into \eq{entropy-eq} written
in the form $\theta\DT\eta=\xi-{\rm div}\,\jj-\theta\eta{\rm div}\,\vv$,
we obtain 
\begin{align}\nonumber
  &\hspace*{-2em}c
  \DT\theta
=\xi
+\theta\frac{\psi_{\FF\theta}''(\FF,\mm,\theta)\!}{\det\FF}{:}\DT\FF
+\theta\frac{\psi_{\mm\theta}''(\FF,\mm,\theta)\!}{\det\FF}
{\cdot}\DT\mm-{\rm div}\,\jj
\\&\text{ with the heat capacity }\
c
=-\theta\,\frac{\!\psi_{\theta\theta}''(\FF,\mm,\theta)\!}{\det\FF}\,,
\label{heat-eq+}\end{align}
which can be understood as the {\it heat equation} for the temperature
$\theta$ as an intensive variable.

The referential {\it internal energy} is given by the {\it Gibbs relation}
$\psi+\theta\eta$. In our Eulerian formulation, we will need rather the actual
internal energy, which, because of \eq{stress-entropy}, equals here to 
\begin{align}\nonumber
&\hspace*{-2.5em}\morelinesunder{\frac{\psi-\theta\psi_\theta'}{\det\FF}}{actual}{internal}{energy}\!\!\!\!\!
=\!\!\!\!\!\linesunder{\frac{\psi(\FF,\mm,0)}{\det\FF}
}
{actual stored}{energy}
\!\!\!\!\!
+\!\!\!\!\!\!
\linesunder{\frac{\COUPLING(\FF,\mm,\theta){-}\theta\COUPLING_\theta'(\FF,\mm,\theta)}
{\det\FF}}{$=:\omega(\FF,\mm,\theta)\:$ thermal part of}{the internal energy} 
\\&\text{with }\ \COUPLING(\FF,\mm,\theta)=\psi(\FF,\mm,\FF)
-\psi(\FF,\mm,0)\,.\label{def-of-omega}\end{align}
In terms of the thermal part of the internal energy $\W=\omega(\FF,\mm,\theta)$
as an extensive variable, the heat equation \eq{heat-eq+} can be written in
the so-called {\it enthalpy formulation}: 
\begin{align}\nonumber
&\hspace*{-2em}\pdt\W+{\rm div}(\vv\W)=\xi
-{\rm div}\,\jj+\frac{\COUPLING'_{\FF}(\FF,\mm,\theta)\!}{\det\FF}{:}\DT\FF
\\[-.4em]&
+\frac{\COUPLING'_{\mm}(\FF,\mm,\theta)\!}{\det\FF}{\cdot}\DT\mm
\hspace{1em}
\text{ with }\ \ \W=\omega(\FF,\mm,\theta)
\,.
\label{Euler-thermodynam3-}\end{align}

\subsection{Thermodynamically coupled system}

The overall system then merges the momentum equation
\eq{Euler-thermodynam1-}, the hysteretic Gilbert equation \eq{LL-friction}
with the Poisson equation \eq{u-eq} for the demagnetizing field,
the heat equation \eq{Euler-thermodynam3-}
together with the kinematic equation \eq{ultimate} and the usual
continuity equation for mass density $\varrho$ transported as an extensive
variable, cf.\ \eq{Euler-thermodynam0} below. We consider a specific
dissipation potential 
\begin{align}
  \hspace*{-1em}\zeta(\theta;\ee,\rr)=\frac12\bbD\ee{:}\ee
  +\frac12\alpha|\rr|^2+\HC|\rr|
\end{align}
with $\bbD$ a 4th-order symmetric tensor of elastic moduli.
Altogether, we arrive to a system of six equations
for $(\varrho,\vv,\FF,\mm,u,\theta)$:
\begin{subequations}\label{Euler-thermodynam}
\begin{align}\label{Euler-thermodynam0}
&\hspace*{-2em}\pdt\varrho=-\,{\rm div}(\varrho\vv)\,,
\\\nonumber
&\hspace*{-2em}\pdt{}(\varrho\vv)={\rm div}\Big(\TT+\KK+\SS-{\rm div}\mathscr{S}
  +\DIS\ee(\vv)-\varrho\vv{\otimes}\vv\Big)
\\[-.5em]\nonumber
    &\hspace*{4em}
  +\mu_0(\nabla\hh)^\top\mm-\mu_0\nabla(\hh{\cdot}\mm)+\varrho\GRAVITY
\\[-.1em]\nonumber&\hspace*{-1em}\text{with }\ \SS={\rm skw}\Big(\Big(\mu_0\hh {-}
 \frac{\psi_\mm'(\FF,\mm,\theta)\!}{\det\FF}\Big){\otimes}\mm\Big)\,,
\\[-.1em]\nonumber
    &\hspace*{1em}\mathscr{S}=
\frac{\kappa(\FF)}{\det\FF}{\rm Skw}\big(\mm{\otimes}\nabla\mm),\ \text{ and }\ 
\ \ \\[-.1em]
    &\hspace*{1em}\hh=\hh_{\rm ext}+\nabla u\ \ \text{ and \ $\TT,\,\KK$\ from (\ref{driving-fields}a,b)}\,,\!\!\!\!
    \label{Euler-thermodynam1}
    \\[-.2em]
&\hspace*{-2em}\pdt\FF=(\Nabla\vv)\FF-(\vv{\cdot}\nabla)\FF\,,
\label{Euler-thermodynam2}
\\[-.2em]&\nonumber
\hspace*{-2em}
\alpha\ZJ\mm+\HC(\theta){\rm Dir}(\ZJ\mm)
\\[-.5em]&\hspace*{-.7em}
-\frac{\mm{\times}\ZJ\mm}{\gamma(\mm,\theta)}
\ni\mu_0\hh-\tt
\ \text{ with $\tt$ from \eq{driving-field}}\,,
\label{Euler-thermodynam4}\\&\label{Euler-thermodynam5}
\hspace*{-2em}u(\xx)=\frac1{4\uppi}\int_\varOmega\frac{(\widetilde\xx{-}\xx){\cdot}\mm(\widetilde\xx)}
{|\widetilde\xx{-}\xx|^3}\,\d\widetilde\xx\ \ \ \text{ for }\ \xx\in\varOmega\,,
\\&\nonumber\hspace*{-2em}
\pdt{\W}
=\xi(\FF,\theta;\ee(\vv)
,\ZJ\mm)
+\frac{\COUPLING'_{\FF}(\FF,\mm,\theta)\FF^\top\!\!}{\det\FF}{:}\ee(\vv)
\\[-.1em]&\nonumber\hspace{7em}
+\frac{\COUPLING_\mm'(\FF,\mm,\theta)}{\det\FF}{\cdot}\DT\mm
-{\rm div}\big(\jj{+}\W\vv\big)
\\&\nonumber\hspace{0em}
\text{ with }\
\xi(\FF,\theta;\ee,
\rr)=\DIS\ee{:}\ee
+\mu_0\alpha|\rr|^2+\mu_0\HC(\theta)|\rr|
\\&\nonumber\hspace{0em}\text{ and }\ \jj=-\COND(\FF,\theta)\nabla\theta\,,
\\[-.9em]&\nonumber\hspace{0em}\text{ and }\ \W=\OMEGA(\FF,\mm,\theta)
=\frac{\COUPLING(\FF,\mm,\theta)-\theta\COUPLING_\theta'(\FF,\mm,\theta)}{\det\FF}
\\[-.2em]&\hspace{0em}\text{ with }\ 
\COUPLING(\FF,\mm,\theta)=\psi(\FF,\mm,\theta)-\psi(\FF,\mm,0).
\label{Euler-thermodynam3}\end{align}\end{subequations}
The skew-symmetric stress $\SS$ and couple-like hyperstress $\mathscr{S}$ come
from the magnetic dipoles and from the exchange energy and
balances the energetics, cf.\ Sect.~\ref{sec-engr} below and the
calculations in \cite{Roub23LTFP}; for a similar skew-symmetric stress
$\SS$ see 
\cite{DeSPoG95ISIS,DeSPoG96CTDF} while the skew-symmetric hyperstress
$\mathscr{S}$ is like in the Cosserat theory by \cite{Toup64TECS}.
The skew-symmetric couple-like hyperstress $\mathscr{S}$ balances the
energetic as \eq{energy+}--\eq{thermodynamic-Euler-engr} below, being
related\footnote{Cf.\ the calculations
\cite[Formulas (2.46)--(2.47)]{Roub23LTFP}.} with the corotation derivative
$\ZJ\mm$ used in \eq{Euler-thermodynam4} and would not be visible if only
the convective
derivative\footnote{Cf.\ e.g.\ \cite{BFLS18EWSE,GKMS22SWPS,KaKoSc21MAWS}.
}
$\DT\mm$ were used there.

\section{Creep and melting-solidification transition}\label{sec-creep}

The solid-type viscoelastic rheology in Section~\ref{sec-elast-mag}
has applications rather in metallic magnets but some other
materials need rather more advanced rheologies, specifically
some a fluidic-type rheology in the shear part of the model. This
concerns, in particular, the geophysical application in Sect.\,\ref{sec-rocks}
where very large displacements occur within long geological periods.

\subsection{Creep in multiplicative decomposition}

Our treatment of finite-strain inelasticity (here creep)
will be based on the \cite{Kron60AKVE} and \cite{LeeLiu67FSEP}
{\it multiplicative decomposition}, as routinely used in plasticity, i.e.
\begin{align}\nonumber\\[-2.7em]\label{Green-Naghdi}
\FF=\Fe\Fp\,,
\end{align}
where $\Fp$ is a {\em inelastic distortion} tensor. This
tensor $\Fp$ is interpreted as a transformation
of the reference configuration into an intermediate stress-free
configuration, which is then mapped into the current configuration by the
{\it elastic strain}\index{elastic strain} $\Fe$.
It is customary to introduce an {\it inelastic distortion rate},
denoted by $\Lp$, and by differentiating \eq{Green-Naghdi} in time
and by using $\FF=\nabla\vv$ to write
\begin{align}\nonumber\\[-2.2em]
	\DT\Fe=(\nabla\vv)\Fe-\Fe\Lp\ \ \ \text{ with }\ \Lp=\DT\Fp\Fpp^{-1}\,.
\label{DTFe}\end{align}
It is also customary to consider the inelastic distortion isochoric, i.e.\
$\det\Fp=1$, which reflects the natural attribute that, volumetrically,
there cannot be any creep while, deviatorically, there can be even very
large creep due to shearing --- e.g.\ rocks in long geological time scales
may easily creep by thousands of kilometers. This nonlinear holonomic
constraint on $\Fp$ is equivalent to the linear constraint ${\rm tr}\,\Lp=0$
if the initial condition on $\Fp$ is isochoric; here ``tr'' denotes the trace
of a square matrix.

The kinematic equation \eq{DTFe} is to be accompanied
by a rule governing the evolution of $\Fp$. As generally
expected in the position of internal variables, this is
a parabolic-type equation (i.e.\ inertia-free), specifically 
\begin{align}\nonumber\\[-2.2em]
  \GM\DT\Fp={\rm dev}\big(\Fetop\psi_{\Fe}'(\Fe,\mm,\theta)\big)\Fp
\end{align}
with $\GM=\GM(\theta)$ a (temperature-dependent) Maxwellian
creep modulus while ``dev'' denotes the deviatoric part
of a tensor.\footnote{The deviatoric part of a stress $\SS$ is
  defined as ${\rm dev}\SS=\SS{-}({\rm tr}\SS)\bbI$ so that
  the trace of ${\rm dev}\SS$ is zero.}
In terms of the inelastic creep distortion rate $\Lp$ from \eq{DTFe},
it read as an algebraic relation 
$\GM\Lp={\rm dev}\big(\Fetop\psi_{\Fe}'(\Fe,\mm,\theta)\big)$
with the so-called Mandel stress as the right-hand side.

\subsection{The coupled system}
This is to be incorporated into the system
\eq{Euler-thermodynam} which then reads as an integro-differential-algebraic\footnote{The system \eq{Max-Euler-thermodynam} can be called differential
when replacing the ``algebraic'' equation \eq{Max-Euler-thermodynam7}
by \eq{DTFe} and the integral equation \eq{Max-Euler-thermodynam5} by
the Poisson equation \eq{u-eq}.}
system of seven equations for $(\varrho,\vv,\Fe,\Lp,\mm,u,\theta)$:
\begin{subequations}\label{Max-Euler-thermodynam}
\begin{align}\label{Max-Euler-thermodynam0}\nonumber\\[-2.1em]
&\hspace*{-2em}\pdt\varrho=-\,{\rm div}(\varrho\vv)\,,
\\\nonumber
&\hspace*{-2em}\pdt{}(\varrho\vv)={\rm div}\Big(\TT+\KK+\SS{-}{\rm div}\mathscr{S}
  +\DIS\ee(\vv)-\varrho\vv{\otimes}\vv\Big)
\\[-.3em]\nonumber
    &\hspace*{4em}
  +\mu_0(\nabla\hh)^\top\mm-\mu_0\nabla(\hh{\cdot}\mm)+\varrho\GRAVITY
  \\[-.1em]\nonumber
    &\hspace*{-1em}
\text{with }\ \TT
=\Big(\frac{\
  \psi_{\Fe}'(\Fe,\mm,\theta)
  \!}{\det\Fe}+\frac{|\nabla\mm|^2\kappa'(\Fe)\!}{2\det\Fe}\:\Big)
\Fetop\!,
\\[.1em]\nonumber
    &\hspace*{1em}\KK=
\frac{\kappa(\Fe)}{\det\Fe}\nabla\mm{\otimes}\nabla\mm
\,,
\\[.1em]\nonumber
    &
    \hspace*{1em}\SS={\rm skw}\Big(\Big(\mu_0\hh {-}
 \frac{\psi_\mm'(\Fe,\mm,\theta)\!}{\det\Fe}\Big){\otimes}\mm\Big)\,,
\\[.1em]\nonumber
    &\hspace*{1em}\mathscr{S}=
\frac{\kappa(\Fe)}{\det\Fe}{\rm Skw}\big(\mm{\otimes}\nabla\mm),\ \text{ and }\ 
\ \ \\[.1em]
    &\hspace*{1em}\hh=\hh_{\rm ext}+\nabla u\,,\!\!\!\!
    \label{Max-Euler-thermodynam1}
    \\[-.2em]
&\hspace*{-2em}\pdt\Fe=(\Nabla\vv)\Fe-(\vv{\cdot}\nabla)\Fe-\Fe\Lp\,,
\label{Max-Euler-thermodynam2}
\\[-.6em]&\nonumber
\hspace*{-2em}
\alpha\ZJ\mm+\HC(\theta){\rm Dir}(\ZJ\mm)-\frac{\mm{\times}\ZJ\mm}{\gamma(\mm,\theta)}
\ni\mu_0\hh
\\[-.2em]
&\hspace*{3.5em}
-\frac{\psi_\mm'(\Fe,\mm,\theta)\!}{\det\Fe}
+{\rm div}\Big(\frac{\upkappa(\Fe)}{\det\Fe}\nabla\mm\Big)\,,
\label{Max-Euler-thermodynam4}
\\&\label{Max-Euler-thermodynam7}
\hspace*{-2em}
\GM(\theta)\Lp={\rm dev}\big(\Fetop\psi_{\Fe}'(\Fe,\mm,\theta)\big)\,,
\\&\label{Max-Euler-thermodynam5}
\hspace*{-2em}u(\xx)=\frac1{4\uppi}\int_\varOmega\frac{(\widetilde\xx{-}\xx){\cdot}\mm(\widetilde\xx)}
{|\widetilde\xx{-}\xx|^3}\,\d\widetilde\xx\ \ \ \text{ for }\ \xx\in\varOmega\,,
\\&\nonumber\hspace*{-2em}
\pdt{\W}
=\xi(\Fe,\theta;\ee(\vv)
,\ZJ\mm,\Lp)
+\frac{\COUPLING'_{\Fe}(\Fe,\mm,\theta)\Fetop\!\!}{\det\Fe}{:}\ee(\vv)
\\[-.5em]&\nonumber\hspace{-1.3em}
-\frac{\!\Fetop\COUPLING'_{\Fe}(\Fe,\mm,\theta)\!}{\det\Fe}{:}\LL
+\frac{\COUPLING_\mm'(\Fe,\mm,\theta)}{\det\Fe}{\cdot}\DT\mm
-{\rm div}\big(\jj{+}\W\vv\big)
\\&\nonumber\hspace{-1em}
\text{ with }\
\xi(\Fe,\theta;\ee,
\rr,\Lp)=\DIS\ee{:}\ee
+\mu_0\alpha|\rr|^2
\\[-.1em]&\nonumber\hspace{8.3em}+\mu_0\HC(\theta)|\rr|+\GM(\theta)|\Lp|^2
\\&
\hspace{-1em}\text{ and }\ \jj=-\COND(\Fe,\theta)\nabla\theta\,,
\text{ and }\ \W=\OMEGA(\Fe,\mm,\theta)\,,
\label{Max-Euler-thermodynam3}\end{align}\end{subequations}
where $\COUPLING$ and $\OMEGA$ are from \eq{Euler-thermodynam3}.

\begin{example}[{\sl Neo-Hookean material}]\upshape\label{exa-1}
For illustration of the structure of the model,
let us consider the data
\begin{subequations}\label{exa}\begin{align}
    &\hspace{-2em}\nonumber
    \psi(\Fe,\mm,\theta)=\frac12K_\text{\sc e}^{}\big(J{-}1\big)^2\!
+ G_\text{\sc e}^{}\big(J^{-2/3}{\rm tr}(\Fe\Fetop){-}3\big)
\\[-.4em]&\hspace{3.em}\label{neo-Hookean2}
+a|\mm|^4\!+b(\theta{-}\theta_\text{\sc c}^{})|\mm|^2\!
+c\theta(1{-}{\rm ln}\theta)\,,
\\&\nonumber\hspace{-2em}\zeta(\theta;\ee(\vv),\ZJ\mm,\Lp)=\frac12K_\text{\sc v}|{\rm div}\,\vv|^2
+G_\text{\sc v}|{\rm dev}\,\ee(\vv)|^2+\alpha|\ZJ\mm|^2
  \\[-.4em]&\hspace{7.em}
+\frac12\GM(\theta)|\Lp|^2+\HC(\theta)|\ZJ\mm|\,
\label{neo-Hookean1}\end{align}\end{subequations}
with $J=\det\Fe$, the elastic bulk and shear moduli $K_\text{\sc e}^{}$
and $G_\text{\sc e}^{}$ and the Stokes viscosity bulk and shear
moduli $K_\text{\sc v}^{}$ and $G_\text{\sc v}^{}$, respectively. 
the heat capacity $c$, and $\theta_\text{\sc c}^{}$ the Curie temperature.
Eventually, $\GM$ stands for the temperature-dependent Maxwellian creep
modulus. Noteworthy, $\psi(\cdot,\mm,\theta)$ is frame indifferent.
Having the isochoric inelastic strain $\Fp$, the model separates the
volumetric (spherical) and the deviatoric parts so that it combines
the Kelvin-Voigt rheology in the volumetric part with the 
Jeffreys rheology, cf.~Figure~\ref{fig2}.
\begin{figure}[]
	\centering
	\includegraphics[width=.46\textwidth]{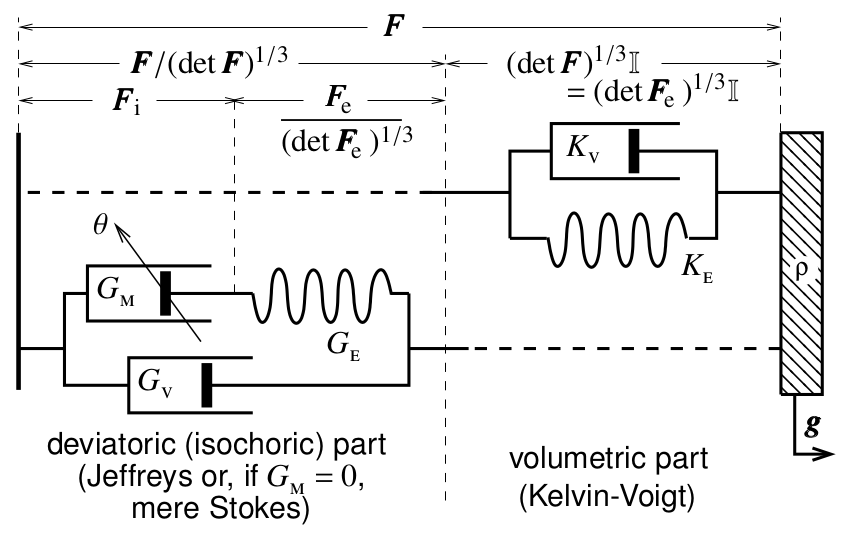}
\caption{A schematic 0-dimensional diagramme of the mixed rheology acting
  differently on volumetric and the deviatoric parts: the Jeffreys rheology
  in the deviatoric part combines Stokes' and Maxwell rheologies in parallel
  and may degenerate to mere Stokes-type fluid if $G_\text{\sc m}^{}$
  vanishes within melting.}\label{fig2}
\end{figure}
\end{example}

\begin{remark}\upshape
Noteworthy, the free energy \eq{neo-Hookean2} is meant referential
while the actual free energy is $\psi(\Fe,\mm,\theta)/\!\det\Fe$.
The corresponding saturation magnetization $m_\text{\sc s}$ in zero external
magnetic field as the magnitude of the minimizers of $\psi(\Fe,\cdot,\theta)$
or, equally, of $\psi(\Fe,\cdot,\theta)/\!\det\Fe$ is given by \eq{ms}
and is independent of $\Fe$. Yet, in compressed ferromagnetic materials,
the saturation magnetization as well as the Curie temperature depends on
hydrostatic (or here lithostatic) pressure, cf.\ e.g.\
\cite{Guga58CSMH,KouWil61MINA,Stac56BFSC,VKMC21PILI}. This can be modelled
by making $b$ in \eq{neo-Hookean2} dependent on $\det\Fe$, or even more
generally replacing $b(\theta{-}\theta_\text{\sc c}^{})$  in \eq{neo-Hookean2}
by $(b_1\theta{-}b_2)$ with $b_1=b_1(\det\Fe)$ and $b_2=b_2(\det\Fe)$.
Such generalization would not contribute to the heat capacity but would
contribute to the Cauchy stress by a pressure $b_1'(\det\Fe)|\mm|^2$.
\end{remark}

\subsection{Energetics of the coupled system}\label{sec-engr}

To reveal the energetics behind the system \eq{Max-Euler-thermodynam}
and in the particular case also behind \eq{Euler-thermodynam}, both considered
  on a fixed domain $\varOmega\subset\R^3$, 
  needs specification of some boundary conditions on the boundary $\varGamma$
  of this domain. For simplicity, let us fix it mechanically
and isolate thermally by prescribing:
\begin{align}\label{BC}
  \hspace{-1em}\vv=\boldsymbol0\,,\ \ \
  \nabla\mm{\cdot}\nn=\boldsymbol0\,,\ \text{ and }\ \:
  \nabla\theta{\cdot}\nn=0\ \ \ \text{ on }\ \varGamma\,,
\end{align}
where $\nn$ denotes the normal to $\varGamma$.
The first condition can be modified to a Navier-type
condition but, minimally, the normal velocity $\vv{\cdot}\nn$ is
to be zero to fix the domain $\varOmega$, otherwise the Eulerian
formulation becomes very cumbersome. When an evolving domain
is needed, some fictitious large fixed domain around the
magnetoelastic material filled with ``air'' is used, being
called a ``sticky-air'' approach in geophysical modelling.
Let us remark that it is used also in engineering, where it is known
rather as a fictitious-domain approach or as an immersed-boundary method.

The {\it energy-dissipation balance} can then be seen by testing
the momentum equation \eq{Max-Euler-thermodynam1} by $\vv$ and integrating
it over $\varOmega$ while using the Green formula, the boundary
conditions, and the continuity equation  \eq{Max-Euler-thermodynam0}
tested by $|\vv|^2/2$ and the flow rule \eq{Max-Euler-thermodynam2}
merged also with \eq{Max-Euler-thermodynam7}. Further, one uses the
Gilbert equation \eq{Max-Euler-thermodynam4} tested by $\ZJ\mm$. The
resulting calculations are quite demanding and we refer the reader
for them to \cite{Roub23LTFP} and, for combination with the creep, also to
\cite{RouTom123IFST}. The resulting balance is
\begin{align}\nonumber
  &\hspace{-2em}\frac{\d}{\d t}\bigg(
  \int_\varOmega\!\!\!\!\!\linesunder{\frac\varrho2|\vv|^2}{kinetic}{energy}\hspace{-1.3em}+\hspace{-1.0em}\linesunder{\frac{\psi(\Fe,\mm,0)\!\!}{\det\Fe}}{stored}{energy}\hspace{-1.0em}+\hspace{-1.0em}\linesunder{\frac{\upkappa(\Fe)|\Nabla\mm|^2\!\!}{2\det\Fe}}{exchange}{energy}\hspace{-1.0em}
    -\hspace{-1.2em}\linesunder{\mu_0\hh
      {\cdot}\mm_{_{_{_{}}}}\!\!\!}{Zeeman}{energy}\hspace{-.7em}\d\xx
    \\&\nonumber\hspace*{-.5em}
+\int_{\R^3}\!\!\!\!\!\!\!\!\!\!\linesunder{\frac{\mu_0}2|\Nabla u|^2\!}{energy of de-}{magnetizing field}\!\!\!\!\!\!\!\!\!\,\d\xx\bigg)
+\int_\varOmega\!\!\!\!\!\linesunder{\xi\big(\Fe,\theta;\ee(\vv),\ZJ\mm,\Lp\big)
    _{_{}}\!\!}{dissipation rate}{from \eqref{Max-Euler-thermodynam3}}\!\!\!
    \d\xx=
 \\&\nonumber\hspace*{-2em}
=\int_\varOmega\!\bigg(\!\!\!\!\linesunder{\!\!\varrho\,\GRAVITY{\cdot}\vv_{_{_{_{}}}}\!\!}{power of}{gravity field}\!\!\!\!\!\!
-\!\!\!\!\!\linesunder{\mu_0\frac{\!\partial\hh_\text{\rm ext}\!}{\partial t}{\cdot}\mm}
{power of}{external field}\!\!\!\!
  \\&
  \hspace*{-.2em}\lineunder{-\,
    \frac{\COUPLING_{\Fe}'\!\!(\Fe,\mm,\theta)
    }{\det\Fe}{:}\DT\Fe
-\frac{\COUPLING_\mm'(\Fe,\mm,\theta)\!}{\det\Fe}{\cdot}\DT\mm
\!}{adiabatic effects}\!\!\!\!\!\bigg)\,\d\xx.\hspace*{-1em}
\label{energy+}
\end{align}

When we add \eq{Euler-thermodynam3} tested by 1, the adiabatic and the
dissipative heat sources cancel with those in \eq{energy+}. Thus we obtain the
{\it total energy balance}
\begin{align}\nonumber
  &\hspace{-2em}\frac{\d}{\d t}\bigg(\int_\varOmega\!\!\!\!\!
  \linesunder{\frac\varrho2|\vv|^2}{kinetic}{energy}\hspace{-1em}+\hspace{-1em}
  \lineunder{\frac{\psi(\Fe,\mm,0)}{\det\Fe}+\OMEGA(\Fe,\mm,\theta)
  }{internal energy}\!\!\!\!
\\[-.2em]&\nonumber\hspace{-1em}
+\!\!\!\!\!\!\linesunder{\frac{\upkappa(\FF)|\Nabla\mm|^2}{2\det\FF}}{exchange}{energy}\hspace{-1em}
  -\hspace{-1em}\linesunder{\mu_0\hh
    {\cdot}\mm_{_{}}\!\!\!}{Zeeman}{energy}\!\!\!\!\,\d\xx
  +\int_{\R^3}\hspace{-1,9em}\linesunder{\frac{\mu_0}2|\Nabla u|^2\!\!}{energy of de-}{magnetizing field}\hspace{-1,3em}\d\xx\bigg)
  \\[-.2em]&\hspace{6em}
=\int_\varOmega\!\!\!\!\!\morelinesunder{\varrho\GRAVITY{\cdot}\vv}{power of}{gravity}{field}\!\!\!\!\!\!\!\!-\!\!\!\!\!\!\linesunder{\mu_0\frac{\!\partial \hh_\text{\rm ext}\!}{\partial t}{\cdot}\mm
  }{power of}{external field}\!\!\!\!
\d\xx\,,
\label{thermodynamic-Euler-engr}\end{align}
where $\OMEGA(\Fe,\mm,\theta)$ is from \eq{def-of-omega}.
The skew-symmetric stress $\SS$ and hyperstress $\mathscr{S}$
in \eq{Euler-thermodynam1} used also in \eq{Max-Euler-thermodynam1}
were devised just to facilitate this desired energetics, see
\cite[Sect.2.4]{Roub23LTFP}.

\subsection{Remarks to analysis of a multipolar variant}

The rigorous mathematical analysis of the above models seems problematic
and there is a certain agreement that some higher-gradient theories
must be involved in them to pursue this analytical goal.\footnote{For an
alternative but less physically justified approach, we refer to
\cite{BFLS18EWSE,GKMS22SWPS,KaKoSc21MAWS} where a gradient term ${\rm div}
\nabla\FF$ has been added into the kinematic equation \eq{Euler-thermodynam2}.}
 Our approach follows the theory by 
 \cite{FriGur06TBBC}, as already
 considered in the general nonlinear context of so-called {\it multipolar
  fluids} by J.~Ne\v cas and his group, cf.\
 \cite{Neca94TMF,NeNoSi89GSIC,NecRuz92GSIV}, as originally inspired by
 \cite{Toup62EMCS} and \cite{Mind64MSLE}.
 Also, the gradient of the inelastic distortion rate $\Lp$ is to be involved.

 The dissipation potential $\zeta$ extends by the higher-order terms
 $\nu_1|\nabla^2\vv|^p/p$ with $p>3$ and $\nu_2|\nabla\Lp|^2/2$.
 This brings an additional hyper-stress contribution
 to the Cauchy stress, namely ${\rm div}(\nu_1|\nabla^2\vv|^{p-2}\nabla^2\vv)$,
 and an contribution ${\rm div}(\nu_2\nabla\Lp)$
 to the Mandel stress on the right-hand side of
 \eq{Max-Euler-thermodynam7}.
 The dissipation rate $\xi$ in \eq{Max-Euler-thermodynam3}
 expands by $\nu_1|\nabla^2\vv|^p+\nu_2|\nabla\Lp|^2$,
 as well as the boundary conditions should be
 extended appropriately, cf.\ 
\cite{Roub22QHLS,Roub23LTFP,RouTom123IFST}
 for quite nontrivial analytical details.

\section{Paleomagnetism in crustal rocks}\label{sec-rocks}

The above devised model \eq{Max-Euler-thermodynam} with \eq{exa} can directly
be applied to {\it thermoremanent paleomagnetism} in crustal rocks. Although
the seven-equation system \eq{Max-Euler-thermodynam} may seem too complicated,
it should be pointed out that it is a minimal scenario if one wants to cover
the involved thermomechanical and thermomagnetical processes, as was
indicated in Figure~\ref{fig1}.

\begin{figure}[]
	\centering
	\includegraphics[width=.49\textwidth]{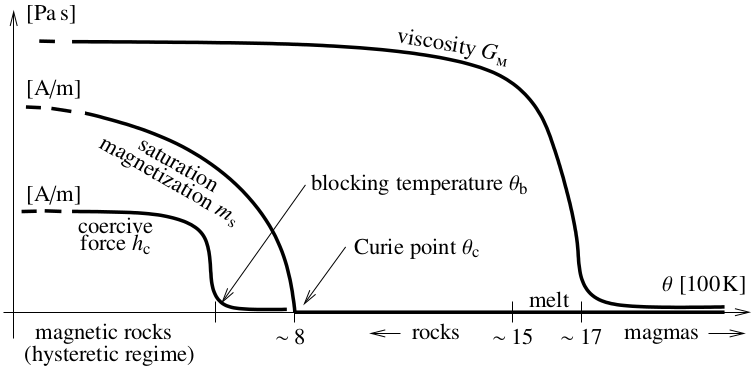}
        \caption{Some temperature-dependent magnetic properties behind
ferri-to-antiferromagnetic phase transition in solid rocks and mechanical
properties behind melting of rocks towards magma.}\label{fig3}
\end{figure}

The modeling is based on a suitable temperature dependence of the
Maxwellian creep modulus $\GM$ in 
\eq{Max-Euler-thermodynam7}, the saturation magnetization $m_\text{\sc s}$
in \eq{ms}, the coercive force $h_{\rm c}^{}$ in \eq{Max-Euler-thermodynam4}.
The mentioned temperature dependence of $m_\text{\sc s}$
can be performed by an appropriate choice of the Curie temperature
$\theta_\text{\sc c}^{}$ in \eq{Landau-ansatz}.
The mentioned temperature dependence of $\GM$ allows for modelling of
the transition between a very fluidic phase (with low $\GM$ of the order
$10^{4\pm3}$Pa\,s) to rather solid rocks (with high $\GM$ of the order
$10^{22\pm2}$Pa\,s). The mentioned temperature dependence of $h_{\rm c}^{}$ allows
for ``freezing'' the magnetization $\mm$ in rocks when they become
sufficiently cold, well below the Curie temperature.
All three mentioned transitions should be properly ordered with
respect to temperature, cf.\ Figure~\ref{fig3}.

The external field $\hh_{\rm ext}$ then plays the role as the geomagnetic
field generated in outer core by magnetodynamical mechanism.

It is important that rocks even with magnetic minerals
undergoing the antiferro-to-ferri magnetic transition
rather than metals undergoing the para-to-ferro magnetic
transition are electrically not conductive. This is why
we could consider only magneto-static approximation
of the full Maxwell electromagnetic system in \eq{u-eq} and
neglected any eddy currents.

The rate-dependent $\alpha$- and $\gamma$-terms in
\eq{Max-Euler-thermodynam3} are actually quite irrelevant  
within slow thermoremanent magnetization and subsequent mechanical
evolution within million-year time scale in crustal
rocks. Yet, they can be relevant within fast processes
in flash magnetization due to strong magnetic fields
as occurs in lightening. This is the mechanism behind the
{\it isothermal remanent magnetization} in cold crustal
rocks.

Besides these two, there is also a {\it viscous remanent magnetization}
which may occur when rocks are exposed in a sufficiently long time by
modern-day magnetic fields which are stronger than geomagnetic field
but anyhow not so strong to lead to an immediate (re)magnetization.
This would need a modification of the linear term $\alpha\ZJ\mm$ in
\eq{Max-Euler-thermodynam4} to a nonlinear, piecewise linear term
distinguishing slow and fast magnetization or, in other words,
a nonquadratic modification of the quadratic term $\alpha|\rr|^2/2$
in the dissipation potential \eq{neo-Hookean2}.

\begin{remark}[{\sl Heterogeneous model}]\label{rem-hetero}\upshape
  Rocks in wider spacial areas are typically substantially heterogeneous,
  as also depicted in Fig.~\ref{fig1}. This can be included in the model,
  beside $\XX$-dependence of the initial conditions, also by allowing for
  an $\XX$-dependence of the data $\psi$, $\DIS$, and $\HC$. In Eulerian
  formulation, $\XX$ serves as a placeholder for $\boldsymbol\xi(\xx)$
  and then the transport equation \eq{transport-xi} for
  $\boldsymbol\xi$ is to be added into the system, cf.\ \cite{RouTom123IFST}
  for a non-magnetic merely thermomechanical variant of this system.
\end{remark}
  
\begin{remark}[{\sl A linearized convective model}]\upshape
  Geophysical modelling at large displacements but small elastic
  strain uses, instead of the deformation gradient $\FF$ and the
  multiplicative decomposition, rather a small strain $\ee(\uu)$ with
  the displacement $\uu=\yy\,{-}\,$identity
  and its Green-Naghdi's additive decomposition to the elastic and the inelastic
  strains expressed in rates using the Zaremba-Jaumann derivative of
  these strains. Such linearization of the model \eq{Max-Euler-thermodynam}
was devised by \cite{Roub23TMPE}.
\end{remark}

\bigskip

\noindent{\bf Acknowledgment.}
Support from the CSF grant no. 23-06220S and the institutional support RVO: 61388998 (\v CR)
is gratefully acknowledged.


\end{document}